\documentclass[a4paper,12pt,doublespace]{article}
\usepackage[utf8x]{inputenc}

\usepackage{dcolumn}

\newcolumntype{.}{D{.}{.}{4}}

\usepackage{setspace}
\usepackage{chngcntr}

\usepackage[left=2cm,top=2cm,right=2cm,bottom=2cm]{geometry}

\usepackage{rotating}
\usepackage{graphicx}
\usepackage{color}

\usepackage{xcolor}
\usepackage{amsmath}
\usepackage{amsfonts}
\usepackage{amssymb}
\usepackage{soul}
\usepackage{bbold}
\usepackage{subfig}
\usepackage{epstopdf}
\usepackage[round]{natbib}

\begin{document}

	\centering
	\Large{Integrating Latent Classes in the Bayesian Shared Parameter Joint Model of Longitudinal and Survival Outcomes}
	
	\vspace{10 mm}
	
	\centering
	\large{Eleni-Rosalina Andrinopoulou$^{1}$, Kazem Nasserinejad$^2$, Rhonda Szczesniak$^{3,4}$ and Dimitris Rizopoulos$^1$
	}
	
	\vspace{10 mm}
	
	\normalsize{
		1. Department of Biostatistics, Erasmus MC, Rotterdam, The Netherlands\\
		2. Department of Hematology, Erasmus MC, Rotterdam, The Netherlands\\
		3. Division of Biostatistics \& Epidemiology and Division of Pulmonary Medicine, Cincinnati Children’s Hospital Medical Center, Cincinnati, United States\\
		4. Department of Pediatrics, University of Cincinnati, Cincinnati, United States
	}
	\vspace{10 mm}
	
	\normalsize{Corresponding author: Eleni-Rosalina Andrinopoulou, Department of Biostatistics, Erasmus MC, PO Box 2040, 3000 CA Rotterdam, The Netherlands\\
		email: e.andrinopoulou@erasmusmc.nl, Tel: +31/10/7043731, Fax: +31/10/7043014}
	
	\singlespacing
	\abstract{
		Cystic fibrosis is a chronic lung disease which requires frequent patient monitoring to maintain lung function over time and minimize onset of acute respiratory events known as pulmonary exacerbations. It is important to characterize the association between key biomarkers such as $FEV_1$ and time-to first exacerbation. Progression of the disease is heterogeneous, yielding different sub-groups in the population exhibiting distinct longitudinal profiles. It is desirable to categorize these unobserved sub-groups according to their distinctive trajectories. Accounting for these latent classes, in other words heterogeneity, will lead to improved estimates of association arising from the joint longitudinal-survival model. 
		
		The joint model of longitudinal and survival data constitutes a popular framework to analyze longitudinal and survival outcomes simultaneously. Two paradigms within this framework are the shared parameter joint models and the joint latent class models. The former paradigm allows one to quantify the strength of the association between the longitudinal and survival outcomes but does not allow for latent sub-populations. The latter paradigm explicitly postulates the existence of sub-populations but does not directly quantify the strength of the association.  
		
		We propose to integrate latent classes in the shared parameter joint model in a fully Bayesian approach, which allows us to investigate the association between $FEV_1$ and time-to first exacerbation within each latent class. We, furthermore, focus on the selection of the optimal number of latent classes.  
	}
	
	\newpage
	
	\doublespacing
	\section{Introduction}\label{Intro}
	
	Cystic fibrosis (CF) is a lethal genetic disorder that primarily affects the lungs. The clinical course of CF is marked by progressive loss of lung function and typically results in respiratory failure. Forced expiratory volume in 1 second (hereafter, $FEV_1$) is the most important clinical indicator in monitoring lung function decline in patients with CF. Patients during follow-up might experience acute respiratory events referred to as pulmonary exacerbations. It is, therefore, of clinical interest to characterize the association between the longitudinal outcome $FEV_1$ and time-to first exacerbation. The motivation for our research comes from the US CF Foundation Patient Registry that consists of patients that were monitored from 2003 until 2015. In particular, we examined a subset of the Registry which consists of 1016 patients. These patients were six years and older and were observed with a median number of follow-up visits equal to six (with a range of 1-93 visits). The average age at baseline is 15 years (with a range of 6-21).  
	
	Several authors have studied the evolution of lung function over time, as summarized in a recent review \citep{szczesniak2017use}, however, to our knowledge little work has been done regarding the association of the lung function such as $FEV_1$ with time-to-event outcomes. In particular, joint modeling of longitudinal $FEV_1$ and survival outcomes in CF was introduced several years ago \citep{schluchter2002jointly}, but has not been further used in CF epidemiology due to the computational burden of this approach. Furthermore, it is well recognized that different unobserved sub-groups of the biomarker $FEV_1$ exhibit different longitudinal profiles \citep{szczesniak2017phenotypes}. Patients can be categorized in several sub-groups (latent classes) with different trajectories. It is, therefore, of high clinical interest to measure the strength of association between $FEV_1$ with the risk of first exacerbation accounting for the latent trajectories. 
	
	The joint model of longitudinal and survival data constitutes a popular framework to analyze longitudinal and survival outcomes jointly \citep{tsiatis2004joint, hickey2016joint}. In particular, two paradigms within this framework are the shared parameter joint models and the joint latent class models. The former paradigm links the longitudinal and the survival process via the random effects \citep{faucett1996simultaneously, wulfsohn1997joint, brown2003bayesian, rizopoulos2011bayesian, rizopoulos2012joint,andrinopoulou2014joint}, which does not allow for latent classes. The latter paradigm \citep{lin2002latent, proust2014joint, rouanet2016joint}, which associates the two processes through latent classes, explicitly postulates the existence of sub-populations but does not directly quantify the strength of the association. 
	
	The aim of the paper is twofold. Firstly, to model the relationship between $FEV_1$ and time-to first exacerbation. For this purpose, we propose a Bayesian shared parameter joint model that integrates latent classes inherent in this heterogeneous population. This model will assess the strength of association between the two outcomes while allowing for latent classes. Secondly, to address a problem that arises in latent class models, which is the selection of the optimal number of classes. Several approaches have been proposed in the literature both in frequentist and Bayesian frameworks, including among others the use of information criterion, Bayes factors and reversible jump MCMC. These approaches are computationally intensive and can require the fit of several models with different numbers of classes, which can be time-consuming. To overcome this problem, we will implement the method of \cite{nasserinejad2017comparison} to our joint model. This method is a pragmatic extension of \cite{rousseau2011asymptotic} criterion that showed that when we overfit a mixture model by assuming more latent classes than present in the data, the superfluous latent classes will asymptotically become empty if the Dirichlet prior on the class proportions is sufficiently uninformative. \cite{nasserinejad2017comparison} performed an extensive simulation study to further investigate this approach and used it as a criterion also in longitudinal studies for obtaining the optimal number of classes by simply excluding latent classes that are negligible in proportion.

	\section{Joint Model Estimation}\label{JM}
	\subsection{Longitudinal submodel}
	
	To account for the fact that the population is heterogeneous and consists of $G$ possible unobserved sub-groups, we postulate a latent class mixed-effects model \citep{verbeke1996linear,proust2005estimation,proust2013analysis}. We let $\boldsymbol{y_{i}}$ denote the longitudinal response vector for the $i$th patient ($i = 1, \dots, n$) obtained at different time points $t_{ij}>0$, $(j = 1, \ldots, n_i)$. In particular, we have
	\begin{equation} \label{MM}
	y_{i}(t \mid v_i = g) = \eta_{ig}(t) + \epsilon_i(t) = \boldsymbol{x}^\top_{i}(t)\boldsymbol{\beta}_g + \boldsymbol{z}^\top_{i}(t)\boldsymbol{b}_{ig} + \epsilon_i(t),
	\end{equation}
	where $v_i = g \ (g = 1, \ldots, G)$ presents the latent class indicator, $\boldsymbol{x}_{i}(t)$ denotes the design vector for the fixed effects regression coefficients $\boldsymbol{\beta}_g$ and $\boldsymbol{z}_{i}(t)$ the design vector for the random effects $\boldsymbol{b}_{ig}$. Moreover, $\epsilon_i(t) \sim N(0, \sigma_y^2)$. For the corresponding random effects, we assume a multivariate normal distribution, namely
	\[
	\boldsymbol{b}_{ig} \sim N(\boldsymbol{0},\boldsymbol{\Sigma}_{bg}),
	\]
	where $N$ denotes the normal distribution and $\Sigma_{bg}$ is the variance-covariance matrix of the random effects. An individual has a probability $\pi_{ig} = P(v_i = g)$ of belonging to latent class $g$. Using a multinomial distribution we obtain the class of each individual as,
	\[
	v_i \sim Multinomial(\pi_{ig}).
	\]
	According to the specification of the latent class mixed-effects submodel (\ref{MM}), both fixed and random effects are class-specific, whereas the measurement error $\epsilon_i(t)$ is not.

	\subsection{Survival submodel}
	We let $\mathrm{\textit{T}}_{i}^*$ denote the true failure time for the $i$-th individual, and $C_i$ the censoring time. Moreover, $T_i=\min(\mathrm{\textit{T}}_{i}^*,C_i)$ denotes the observed failure time and $\delta_i = \{0, 1\}$ is the event indicator where zero corresponds to censoring.
	We postulate a joint model for the relationship between the survival and the longitudinal outcome. Specifically, we have
	\begin{equation}\label{JM}
	h_{i}(t \mid v_i = g)=h_{0g}(t)\exp [\boldsymbol{\gamma}_g^{\top} \boldsymbol{w}_{i}+ \alpha_g \eta_{ig}(t)],
	\end{equation}
	where $\boldsymbol{w}_{i}$ is a vector of baseline covariates with a corresponding vector of regression coefficients $\boldsymbol{\gamma}_g$ and $h_{0g}(t)$ is the baseline hazard. Specifically, the B-splines baseline hazard function is assumed $\log h_{0g}(t)=\gamma_{h_0g,0}+\sum_{q=1}^{Q}\gamma_{h_0g,q}B_q(t,\boldsymbol{\nu})$,
	where $B_q(t,\boldsymbol{\nu})$ denotes the $q$-th basis function of a B-spline with knots $\nu_1,\dots, \nu_Q$ and $\boldsymbol{\gamma}_{h_0g}$ the vector of spline coefficients. The knots are placed at equally spaced percentiles of the observed event times. Furthermore, $\alpha_g$ denotes the association parameter for the $g$th class. According to the specification of the survival submodel (\ref{JM}) the baseline covariates, the baseline hazard and the association parameter are class-specific parameters. The proposed model goes beyond the standard joint model and joint latent class model where a single or no association parameter is assumed and provides a class-specific association. This is a more realistic assumption for the motivating data set since it is clinically expected that the risk of the first exacerbation will be higher when the rate of $FEV_1$ decline is faster. Accounting for these latent classes will lead to improved estimates of association arising from the joint model.

	\section{Bayesian Estimation}
	
	We employ a Bayesian approach where inference is based on the posterior distribution of parameters in the model. We use Markov chain Monte Carlo (MCMC) methods to estimate the parameters of the proposed model. The likelihood of the model is derived under the assumption that the longitudinal and survival processes are independent given the random effects. Moreover, the longitudinal responses of each subject are assumed independent given the random effects \citep{rizopoulos2012joint}. The likelihood contribution for the $i$-th patient is written as
	\[
	\begin{array}{ll}%
	\lefteqn{p(\boldsymbol{y}_i,T_i,\delta_i \mid v_i = g, \boldsymbol{\theta}, \boldsymbol{b}_{ig}) =} \\
	&  \sum_{g=1}^{G} \pi_{ig} \Big\{ \prod_{j=1}^{n_{i}} \bigg[ p(y_{ij}\mid v_i = g, \boldsymbol{\theta}_y, \boldsymbol{b}_{ig}) \bigg] p\{T_i,\delta_i\mid v_i = g, \mathcal \eta_{ig}(T_i),\boldsymbol{\theta}_s, \boldsymbol{b}_{ig}\} \Big\},
	\end{array}
	\]
	
	where $\boldsymbol{\theta} = (\boldsymbol{\theta}_s^{\top}, \boldsymbol{\theta}_{y}^{\top}, \pi_{ig})^{\top}$ with $\boldsymbol{\theta}_y = (\boldsymbol{\beta}_g, \sigma_y, \boldsymbol{\Sigma}_{bg})$ and $\boldsymbol{\theta}_s = (\boldsymbol{\gamma}_g, \alpha_g, \boldsymbol{\gamma}_{h_0g})$. 
	
	The likelihood contribution of the longitudinal outcome takes the form
	\begin{eqnarray*}
		p(y_{ij}\mid v_i = g, \boldsymbol{\theta}_{y}, \boldsymbol{b}_{ig})=(2\pi \sigma_y)^{-1/2} \exp{\biggl[  -\frac{(y_{ij} -\boldsymbol{x_{ij}}^\top\boldsymbol{\beta}_g - \boldsymbol{z_{ij}}^\top\boldsymbol{b}_{ig})^2}{2\sigma_y^2} \biggr] }.
	\end{eqnarray*} 
	
	The likelihood contribution of the survival model is given by
	\begin{eqnarray*}
		\lefteqn{p\{T_i,\delta_i\mid v_i = g, \mathcal \eta_{ig}(T_i),\boldsymbol{\theta}_s , \boldsymbol{b}_{ig}\}= } \\
		&&  \exp\biggr[\gamma_{h_{0}g,0} + \sum_{q=1}^Q\gamma_{h_{0}g,q}B_q(T_i,\boldsymbol{\nu})+ \gamma_g^{\top}\boldsymbol{w}_{i}+ \eta_{ig}(T_i)  \boldsymbol{{\alpha}}_{g} \biggr] ^{I(\delta_i=1)}\times \\
		&& \exp{\biggl\{-\exp{(\boldsymbol{\gamma}_g^{\top} \boldsymbol{w}_{i})\int_0^{T_i}} \exp{\biggl[\gamma_{h_{0}g,0} + \sum_{q=1}^Q \gamma_{h_{0}g,q}B_q(s,\boldsymbol{\nu})+ \eta_{ig}(s) \boldsymbol{\alpha}_{g},  \biggr]}ds  \biggr\}}.
	\end{eqnarray*} 
	
	The posterior distribution is written as
	\[
	\begin{array}{ll}%
	p(\boldsymbol{\theta}, \boldsymbol{b}_{g} \mid \boldsymbol{y},\boldsymbol{T},\boldsymbol{\delta}) = 
	\prod_{i=1}^{n}  p(\boldsymbol{y}_i,T_i,\delta_i \mid v_i = g, \boldsymbol{\theta}, \boldsymbol{b}_{ig})p(\boldsymbol{b}_{ig}\mid v_i = g, \boldsymbol{\theta}_{y})p(\boldsymbol{\theta}).
	\end{array}
	\]
	where
	\[
	p(\boldsymbol{b}_{ig}\mid v_i = g, \boldsymbol{\theta}_{y}) = [2\pi\det(\boldsymbol{\Sigma}_{bg})]^{-1/2}\exp{\biggr(- \frac{\boldsymbol{b}_{ig}^\top \boldsymbol{\Sigma}_{bg}^{-1} \boldsymbol{b}_{ig}}{2} \biggr) },
	\]
	and $p(\boldsymbol{\theta})$ denotes the prior distributions.
	
	A commonly used prior in mixture models for the class probability is a Dirichlet distribution. In particular,
	\[
	\pi_{ig} = P(v_i = g) \sim Dirichlet(\boldsymbol{a}).
	\]
	Small values of $\boldsymbol{a} = \{a_1\dots a_G\}$ correspond to a less informative prior and a flat prior distribution is obtained when each $a_g$ is equal to $1$. The selection of $\boldsymbol{a}$ is an important task and will be discussed in Section~\ref{NumClass}. Standard priors can be assumed for the rest of the parameters. In particular, for the coefficients of the longitudinal fixed effects, the survival covariates and the baseline hazard, normal priors can be taken. For the variance-covariance matrix of the random effects we can assume an inverse Wishart prior, while for the precision parameter of the longitudinal outcome we can assume an gamma prior.

	\subsection{Selection of Number of Classes}\label{NumClass}
	An important task in latent class models is to identify the optimal number of classes. Several approaches have been previously proposed for choosing the optimal number of classes in both frequentist and Bayesian settings. Common examples are the Bayesian information criterion (BIC) \citep{schwarz1978estimating}, deviance information criterion (DIC) \citep{celeux2006deviance} and other Bayesian approaches such as Bayes factor and reversible jump MCMC algorithm \citep{green1995reversible}. A drawback of the aforementioned approaches is that they are computationally intensive and some require the fit of models assuming different numbers of classes, which might be time-consuming for complex models such as the joint models of longitudinal and survival outcomes. 
	
	An interesting alternative was proposed by \cite{rousseau2011asymptotic}, where they proved that in overfitted mixture models (with more latent classes than present in the data), the superfluous latent classes will asymptomatically become empty if the Dirichlet prior on the class proportion is sufficiently uninformative. Recently, \cite{nasserinejad2017comparison} used this approach and proposed a latent class selection procedure for longitudinal models. An overfitted mixture model converged to the true mixture by assigning a small portion of individuals to empty classes, if the parameters of the Dirichlet prior $\boldsymbol{a}$ are smaller than $d/2$, where $d$ is the number of class-specific parameters. Furthermore, uninformative priors for the rest of the parameters are required. The steps are described as follows: 
	
	\begin{itemize}
		\item First, a latent class model with a large enough number of latent classes is fitted. 
		\item Then, the number of non-empty classes at each iteration is calculated as:
		\[
		g_{k,opt} = G - \sum_{g=1}^G I\Bigg(\frac{n_{k,g}}{n} \leq \psi\Bigg),
		\]
		where $G$ is the total number of classes, $k$ represents the iteration, $n_{k,g}$ is the number of patients in class $g$ at iteration $k$, $n$ is the total number of patients and $\psi$ is a predifined value. 
		\item After obtaining the non-empty classes per iteration, the posterior mode of the non-empty classes is calculated. 
		\item Finally, the model with the optimal number of classes which are the non-empty classes is refitted. 
	\end{itemize}
	
	Advantages of this approach are that it is easy to implement even in such complex models and it is not influenced by the label switching problem since we observe the non-empty classes at each iteration. The only time that we need to correct for label switching is when we fit the final model with the optimal number of classes. Furthermore, this approach requires us to fit the model only two times, (namely one with the high number of classes and one with the optimal number of classes) instead of assuming all possible number of classes, therefore decreasing computational burden. It has been shown through extensive simulations in the longitudinal setting that this method performs better than alternative model selection criteria such as BIC and DIC \citep{nasserinejad2017comparison}.

	\section{Analysis of the CF data}\label{Analysis}

	\begin{figure}[h!]
		\centerline{%
			\includegraphics[width = \linewidth]{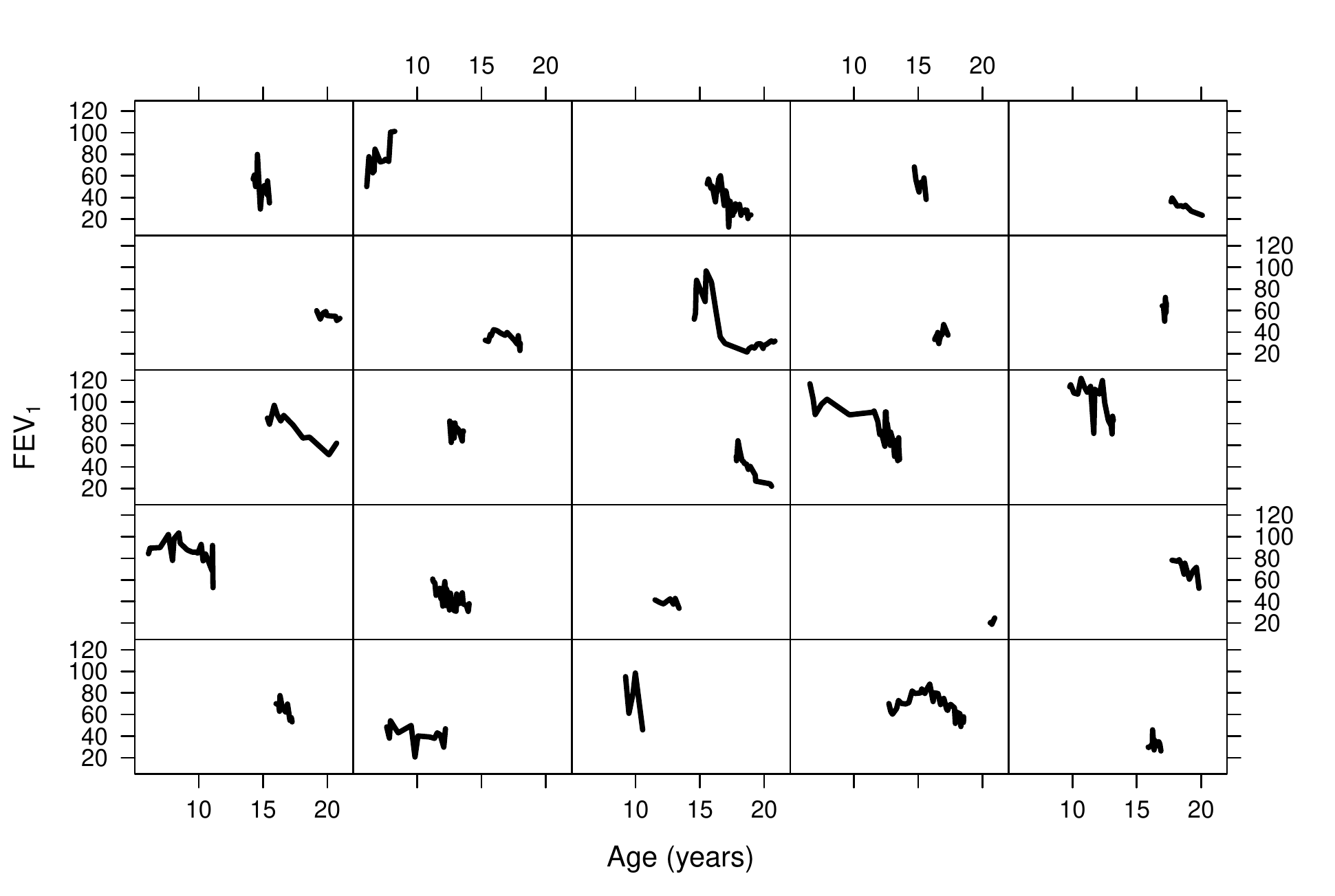}}
		\caption{Individual $FEV_1$ evolutions of 25 randomly selected patients with more than two repeated measurements.}
		\label{indfev1}
	\end{figure}

	In this section we present the analysis of the motivating data set introduced in Section~\ref{Intro}. Our primary focus is to investigate the association between $FEV_1$ and time-to first exacerbation by taking into account that we have sub-groups with different evolution over time for $FEV_1$. The first step is to obtain the optimal number of classes that can explain the heterogeneity of the population. From the literature, it is known that two or three classes are observed for the evolution of $FEV_1$ outcome \citep{szczesniak2017phenotypes}. Therefore, for the selection process, we fitted a joint model assuming six classes. For the longitudinal outcome, we assumed a linear mixed-effects submodel including natural cubic splines for time (modeled as age, in years) with two internal knots at 13.76 and 17.62 years (corresponding to 33.3\% and 66.67\% of the observed follow-up times) in both the fixed and random effects parts. The DIC criterion and subject-specific plots (with observed and predicted values) were used to investigate the need of non-linear evolution over time in a mixed-effects linear model. Furthermore, we corrected for some baseline characteristics. These variables, together with descriptive statistics, are presented in Table~\ref{Desc}. In Figure~\ref{indfev1} the $FEV_1$ evolutions of 25 randomly selected patients with more than two repeated measurements are presented.

	\begin{table}[h!]
		\caption{Descriptive statistics of the variables that were used in the model.}
		\centering
		\begin{tabular}{lr}
			\hline\hline
			& \textbf{Percentage} \\
			\hline\hline
			Gender: & \\
			\ \ Males & 43 \\
			\ \ Females & 57\\
			\hline
			Number of F508del alleles (genotype):\\ 
			\ \ Homozygous&   53\\
			\ \ Heterozygous&   32\\ 
			\ \ Neither  &   6\\
			\ \ Missing & 9\\
			\hline
			Hispanic:  & \\
			\ \ Yes & 8 \\
			\ \ No  & 92\\
			\hline
			White: & \\
			\ \ Yes & 98 \\
			\ \ No & 2 \\
			\hline
			SESlow (Using state/federal or having no \\ insurance is a marker of low socioeconomic status): \\ 
			\ \ Yes& 48 \\ 
			\ \ No & 52 \\
			\hline
			MRSA (Methicillin-resistant Staphylococcus aureus): \\ 
			\ \ Yes & 16 \\ 
			\ \ No & 84 \\
			\hline
			MSSA (Methicillin-sensitive Staphylococcus aureus): \\ 
			\ \ Yes & 21 \\
			\ \ No  & 79 \\
			\hline
			Pa (Pseudomonas aeruginosa): \\ 
			\ \ Yes & 46 \\ 
			\ \ No   & 54 \\
			\hline
			Aspergillus: \\ 
			\ \ Yes & 28 \\
			\ \ No    & 72 \\
			\hline
			CFRD (CF-related diabetes): \\  
			\ \ Normal & 73 \\
			\ \ Impaired & 7\\ 
			\ \ CFRD with or without fasting hyperglycemia&  19 \\
			\hline
			PancEnzymes (Taking a pancreatic enzyme supplement, \\ marks pancreatic insufficiency):  \\ 
			\ \ Yes & 40 \\
			\ \ No & 60 \\
			\hline\hline
			& \textbf{Mean (standard deviation)} \\
			\hline\hline
			Numvisityr \\ (Number of visits at the last follow-up within the prior year) & 5 (3)\\
			\hline\hline
		\end{tabular}
		\label{Desc}
	\end{table}

\clearpage
	
	Specifically, the model takes the form,
	\[
	\begin{array}{ll}%
	y_{i}(t) &= \eta_{ig}(t) + \epsilon_i(t) = \boldsymbol{\beta}_{0g} + \sum_{\omega =1}^3
	\boldsymbol{\beta}_{\omega g} \tt{ns(Age_i, \omega)} +\\
	& \boldsymbol{\beta}_{4g} \tt{Gender} +  \sum_{\omega =5}^7 \boldsymbol{\beta}_{\omega g} \tt{F508} + \boldsymbol{\beta}_{8g} \tt{Hispanic} + \boldsymbol{\beta}_{9g} \tt{White} + \boldsymbol{\beta}_{10g} \tt{SESlow} +  \\
	& \boldsymbol{\beta}_{11g} \tt{MRSA} + \boldsymbol{\beta}_{12g} \tt{MSSA} + \boldsymbol{\beta}_{13g} \tt{Pa} + \boldsymbol{\beta}_{14g} \tt{aspergillus} +  \sum_{\omega =15}^{16} \boldsymbol{\beta}_{\omega  g} \tt{CFRD} + \\
	& \boldsymbol{\beta}_{17g} \tt{PancEnzymes} + \boldsymbol{\beta}_{18g} \tt{numVisityr} + \sum_{\omega =19}^{21}  \boldsymbol{\beta}_{\omega  g} \tt{ns(Age_i, \omega - 18):Gender} + \\
	& \sum_{\omega  =22}^{30} \boldsymbol{\beta}_{\omega  g} \tt{ns(Age_i,\omega - 21):F508} +  \sum_{\omega =31}^{33}  \boldsymbol{\beta}_{\omega  g} \tt{ns(Age_i, \omega - 30):SESlow} + \\
	& \sum_{\omega =1}^3 \boldsymbol{b}_{\omega  g} \tt{ns(Age_i, \omega)} + \epsilon_i(t).
	\end{array}
	\]

	To investigate the association between $FEV_1$ and time-to first exacerbation, we postulated the proposed joint latent class model:
	\[
	h_{i}(t,\boldsymbol{\theta}_s)=h_{0g}(t)\exp [\boldsymbol{\gamma}_g \tt{Gender}_i + \alpha_g \eta_{ig}(t) ].
	\]
	For the baseline hazard we assumed a quadratic B-splines basis with 8 equi-distance internal knots ranging from zero until 19.25 years. 
	
	In the Dirichlet distribution for the prior of the class probability, following the recommendation in \cite{nasserinejad2017comparison}, we assumed $\boldsymbol{a}$ smaller than $d/2$ (where $d$ is the number of class-specific parameters). To ensure that we have the same scale for the coefficients of the covariates in order to easier select uninformative priors, we standardized the $FEV_1$ outcome and the continuous variables (age and numVisityr). Relatively uninformative priors were selected for the parameters in the model. These priors are as follows: 
	\begin{itemize}
		\item $\boldsymbol{\beta_g} \sim N(0, 1000)$,
		\item $\gamma_g \sim N(0, 1000)$,
		\item $\gamma_{h_0g,q} \sim N(0, 1000)$,
		\item $\alpha_g \sim N(0, 100)$,
		\item $\sigma_y^2 \sim GA^{-1}(0.01,0.01)$
		\item $\boldsymbol{\Sigma}_{bg} \sim W^{-1}(M, df)$,
	\end{itemize}
	where $GA^{-1}$ denotes the inverse gamma distribution and $W^{-1}$ denotes the inverse Wishart distribution with $M=diag(0.01)$ being the scale matrix and $df$ the degrees of freedom which is set as the total number of the random effects. For the variance of the association parameter no large variance was required to ensure that we have a uninformative prior since, with the standard joint model we obtained an association parameter smaller than 0.1. The selection of the variances of these priors was investigated with simulations. We ran the MCMC using a single chain with 300,000 iterations, 250,000 burn-in and 10 thinning. The results indicate the presence of three or four classes, assuming that a class is empty if it contains 10 to 15\% of the patients (10\%$\leq \psi \leq$15\%). Since it is established in the literature that two or three classes are present in such populations, we decided to continue with three classes \citep{moss2016comparison, szczesniak2017phenotypes}.
		
\begin{figure}[h!]
		\centerline{%
			\includegraphics[width = \linewidth]{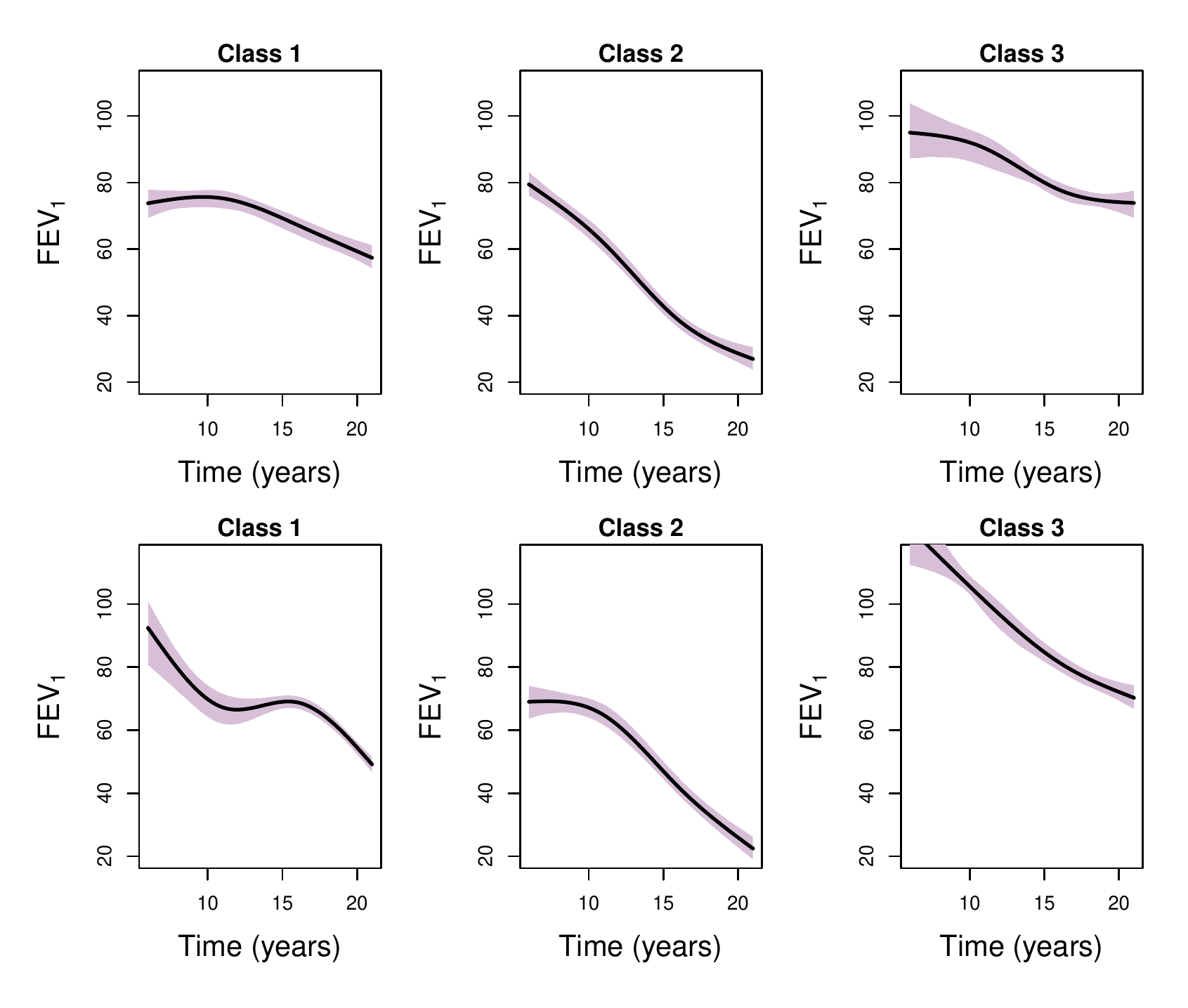}}
		\caption{Evolution of the longitudinal outcome $FEV_1$ per class assuming patients who are F508del homozygotes, non-Hispanic, White, without low SES, are not infections with MRSA, MSSA or aspergillus, do not use pancreatic enzyme, do not have pseudomonas aeruginosa, have normal CFRD and had five visits within the prior year (which is the mean value of all observations). The upper plots represent female patients while the lower plots represent male patients (posterior mean and credible interval).}
		\label{fev1}
	\end{figure}

\begin{figure}[h!]
		\centerline{%
			\includegraphics[width = \linewidth]{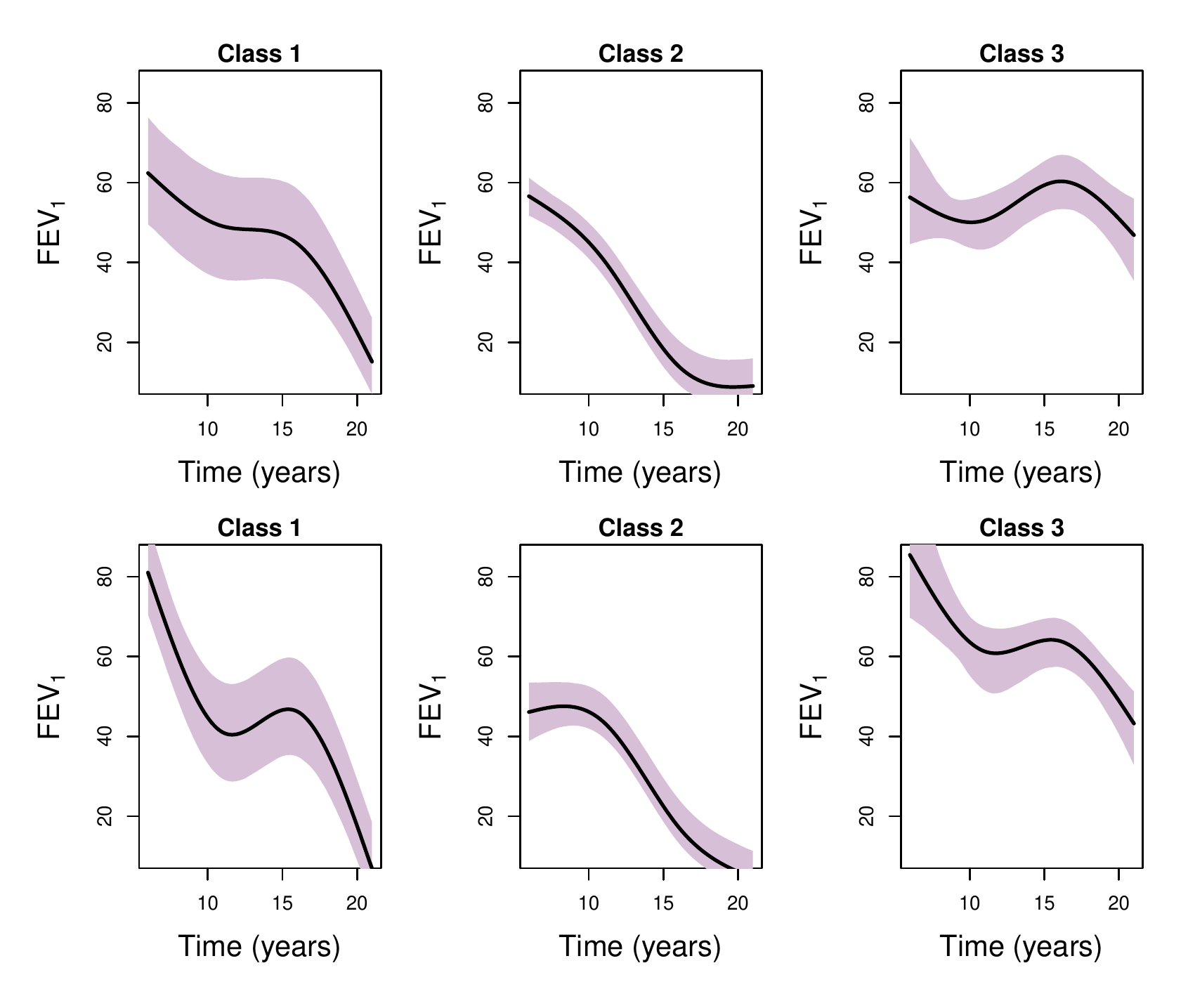}}
		\caption{Evolution of the longitudinal outcome $FEV_1$ per class assuming patients who are F508del homozygotes, Hispanic, White, have low SES, are infections with MRSA, MSSA, aspergillus, use pancreatic enzymes, have pseudomonas aeruginosa, have impaired CFRD and had eight visits within the prior year. The upper plots represent female patients while the lower plots represent male patients (posterior mean and credible interval).}
		\label{fev1b}
	\end{figure}

\begin{figure}[h!]
		\centerline{%
			\includegraphics{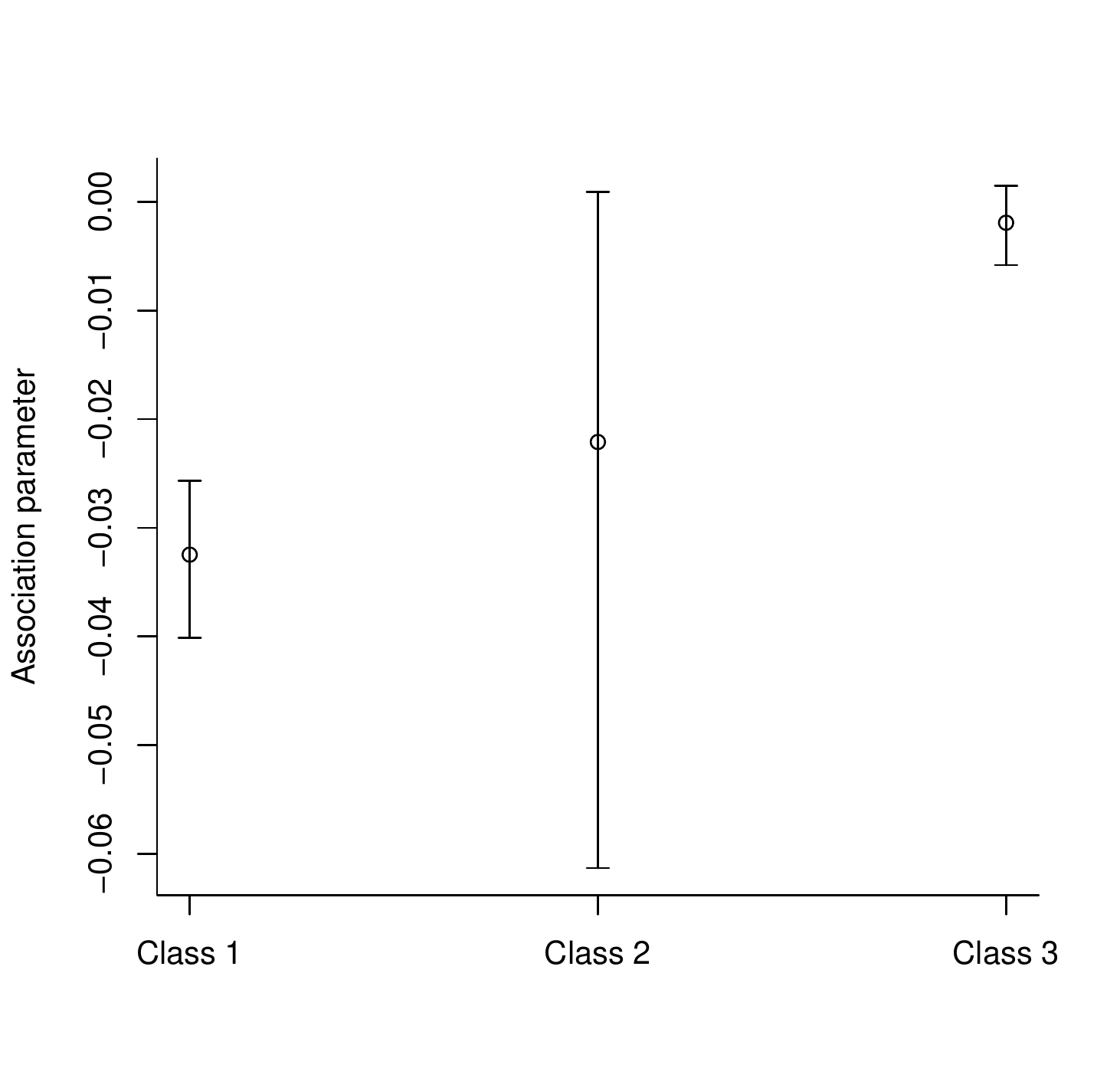}}
		\caption{Mean and credible interval of the association parameter per class.}
		\label{alpha}
	\end{figure}

	We reran the model assuming three classes and the normal scale of the continuous covariate age and $FEV_1$ outcome (we standardized only the numVisityr variable). We ran the MCMCs with a single chain for 500,000 iterations, with a burn-in of 450,000 and thinning of 10 and we fixed the label switching problem. Convergence was monitored by trace plots. Table A1 in the Appendix shows the mean and standard deviation of age (at baseline), $FEV_1$ (at baseline) and number of visits (at last follow-up) per class, while Table A2 shows the percentage of the categorical variables (at baseline) pes class. In Figure~\ref{fev1} we illustrate the evolution of the longitudinal outcome in each class assuming patients who are F508del homozygotes, non-Hispanic, White, without low SES, are not infections with MRSA, MSSA or aspergillus, do not use pancreatic enzyme, do not have pseudomonas aeruginosa, have normal CFRD and had five visits within the prior year (which is the mean value of all observations). In particular, the upper plots represent female patients while the lower plots represent male patients. We obtain a faster progression in class one for both females and males. Patients in class two have a stable evolution in the middle of the follow-up and patients in class three are stable throughout the follow-up period. In addition, patients in class one and two start from a higher $FEV_1$ compared to patients in class three. In Figure~\ref{fev1b} we illustrate the evolution of the longitudinal outcome in each class assuming patients who are F508del homozygotes, Hispanic, White, have low SES, are infections with MRSA, MSSA, aspergillus, use pancreatic enzymes, have pseudomonas aeruginosa, have impaired CFRD and had eight visits within the prior year. Again, the upper plots represent female patients while the lower plots represent male patients. We obtain that patients in these classes start from a lower $FEV_1$ value compared Figure~\ref{fev1}. In addition, we observe a faster progression in class one for both female and male patients. The mean and the credible interval of the MCMC samples of the association parameters per class are presented in Figure~\ref{alpha}. We obtain a weak association between $FEV_1$ and time-to first exacerbation for the second and third class, while a stronger negative association for class one.

\newpage

	\section{Simulations}\label{Sim}
	
	We performed a series of simulations to investigate the proposed class selection method on the joint modeling framework. 
	
	\subsection{Design}
	
\begin{figure}[h!]
		\centerline{%
			\includegraphics[width = \linewidth]{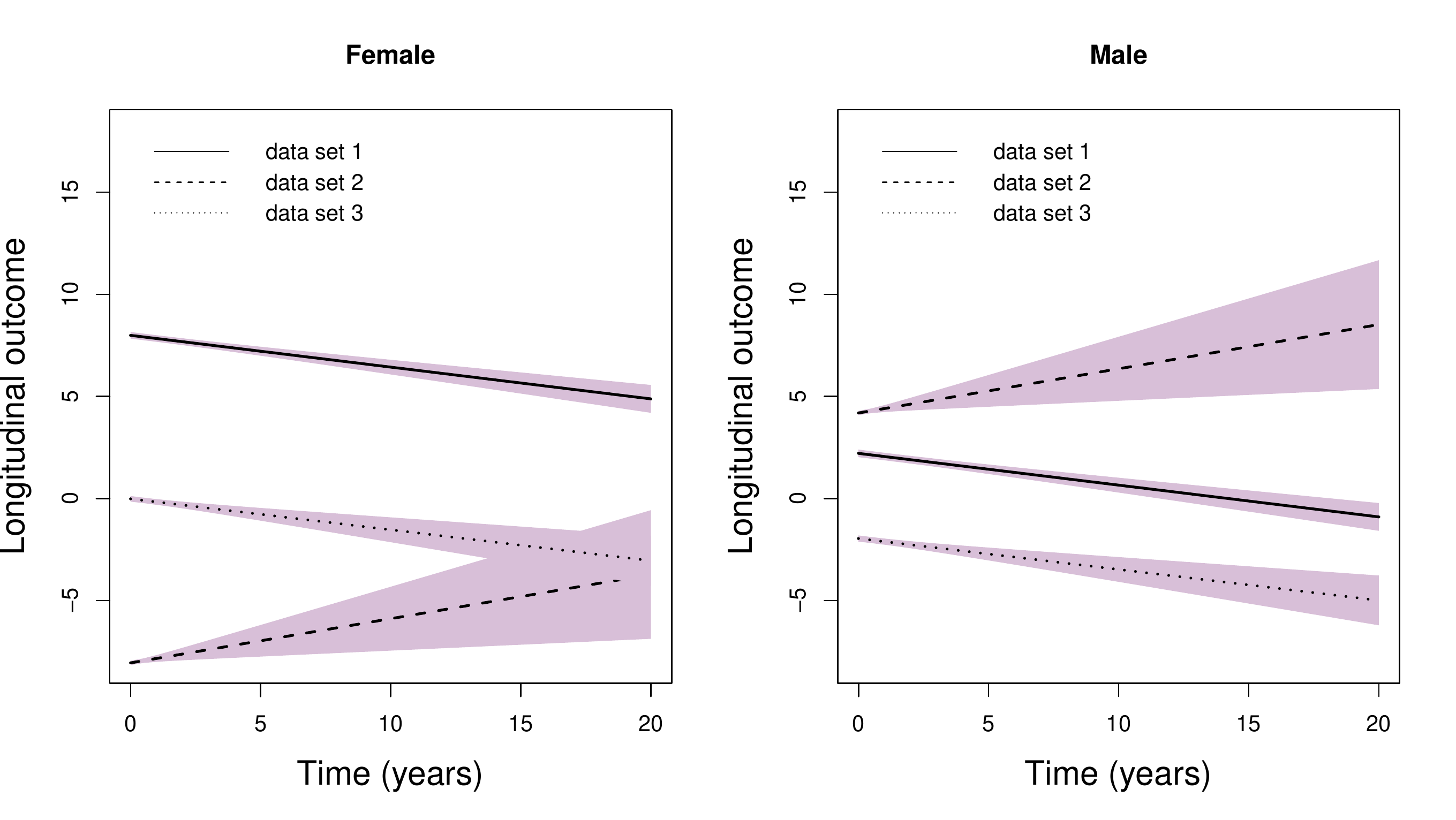}}
		\caption{Evolution of the longitudinal outcome per group from the simulation parameters for each one of the three data sets.}
		\label{SimParamPlots}
	\end{figure}

	We assumed $N_1 = 350$, $N_2 = 525$ and $N_3 = 1050$ patients with maximum number of repeated measurements equal to ten. To simulate the continuous longitudinal outcome, we used the following linear mixed-effects model per data set. In particular,
	\[
	y_{i}(t) = \eta_{i}(t) + \epsilon_{i}(t) = \beta_{0} +  \beta_{1} \tt{male}_i + \beta_{2} t + b_{0i} + b_{1i}t+ \epsilon_{i}(t),
	\]
	where $\epsilon_i \sim N(0,\sigma_y^2)$ and $\boldsymbol{b}_i=(b_{0i}, b_{1i}) \sim N_2(\boldsymbol{0},\boldsymbol{\Sigma}_b)$. For simplicity, we adopted a linear effect of time for both the fixed and the random part, and corrected for a binary variable ($\tt{male}_i$). Time $t$ was simulated from a uniform distribution between zero and $19.5$.
	For the survival part, we assumed the following model:
	\[
	\begin{array}{ll}%
	h_{i}(t)&=h_{0}(t)\exp \Bigg\{\gamma_{}^\top \tt{Age}_i + \alpha \eta_{i}(t) \Bigg\}.
	\end{array}
	\]
	The baseline risk was simulated from a Weibull distribution $h_{0g}(t)=\xi t^{\xi -1}$. For the simulation of the censoring times, an exponential censoring distribution was chosen so that the censoring rate was between 40\% and 60\%. Age was simulated from a normal distribution with mean 45 and standard deviation 15.7.
	\newline
	Under this setting we simulated three different data sets that have different parameters for the fixed effects in the longitudinal submodel, the baseline covariates and baseline hazard in the survival submodel, the variance-covariance matrix of the random effects and the association parameter (more details are presented in Table~\ref{SimParam}). Figure~\ref{SimParamPlots} illustrates the evolution of the longitudinal outcome per group from the simulation parameters for each one of the three data sets.

	\begin{table}[h!]
		\caption{Simulation parameters for the three data sets.}
		\centering
		\begin{tabular}{lrrrrrrr}
			\hline
			& $\boldsymbol{\beta}$ & $\sigma_y$ & $diag\{\boldsymbol{\Sigma}_b\}$ & $\xi$ &$\mu_c$&$\boldsymbol{\gamma}$ & $\alpha$  \\
			\hline
			\textit{Data set 1} & && & & \\
			\hline
			& (Intercept) = 8.03 & 0.69     & 0.87 & 1.8 & 10 & (Intercept) = -4.85  & 0.38\\
			& Male = -5.86 & & 0.02 & && Age = -0.02 &  \\
			& Time = -0.16 &&& & & &\\
			\hline
			\textit{Data set 2}  &&&& & & \\
			\hline
			& (Intercept) = -8.03 & 0.69     &  0.02 & 1.4 & 10 & (Intercept) = -4.85  & 0.08\\
			& Male = 12.20 & & 0.91 & && Age = 0.09 &  \\
			& Time = 0.46 &&& & & &\\
			\hline
			\textit{Data set 3} &&&& & & \\
			\hline
			& (Intercept) = 0.03 & 0.69    & 0.28& 1.8 & 10 & (Intercept) = 2.85  & 0.58\\
			& Male = -1.96 & & 0.31 & && Age = -0.12 &  \\
			& Time = -0.01 &&& & & &\\
			\hline
		\end{tabular}
		\label{SimParam}
	\end{table}

	\subsection{Analysis}
	
	In order to investigate the proposed class selection approach, we applied the model in three different Scenarios. For Scenario I we combined all three data sets assuming $N_1 = 350$ individuals in each of them, for Scenario II we combined the first two data sets with $N_2 = 525$ and finally, for Scenario III we used only the first data set with $N_3 = 1050$. We fitted the proposed joint model assuming six classes where all the parameters were class-specific except for the measurement error in the mixed-effects model. These include the fixed effects from the longitudinal submodel (3 parameters), the baseline covariates from the survival model (2 parameters), the baseline hazard (5 parameters) from the survival submodel, the variance-covariance matrix of the random effects (3 parameters) and the association parameter (1 parameter). To simplify the simulations, for the baseline hazard we assumed quadratic B-splines basis with 4 equally distance internal knots. We assumed ${a_g} = 6.9$ which is smaller than the total number of the class-specific parameters divided by two. The same priors were used as in the application and we ran the MCMC using a single chain with 50,000 iterations, 25,000 burn-in and 10 thinning. 
	
	We performed 150 simulations per Scenario. We compared our proposed method with the joint latent class model using function \textsf{Jointlcmm} from the \textsf{lcmm} package in \textsf{R} developed by \cite{proust2015estimation}, where we used the BIC as criterion. In particular, we assumed that the covariates from the fixed effects in the mixed-effects model, the variance-covariance of the random effects, the baseline covariate and baseline hazard in the survival model are class-specific. We, furthermore, assumed a cubic M-splines baseline risk function. 
	
	\subsection{Results}
	
	The results from the different Scenarios assuming different cut off percentage $\psi$ indicating when a class is defined as empty are illustrated in Table~\ref{SimRes}. In particular, we present the percentage of true number of classes and the mode of the number of classes.
	
	For Scenario I, we obtain the highest percentage when assuming $\psi$ to be between 12-15\%. In particular, we obtain around 55\% of the time the correct number of classes and a mode equal to the correct number of classes (three). On the other hand, the BIC in the joint latent class model selects only 20\% of the time the correct number of classes. Furthermore, this method seems to underestimate the true number of classes (mode equal to one). 
	
	For Scenario II, we obtain the highest percentage when $\psi$ is between 8-15\%. In particular, we obtain around 50\% of the time the correct number of classes and a mode equal to the correct number of classes (two). On the other hand, the BIC in the joint latent class model selects only 12\% of the time the correct number of classes. Similar to Scenario I, this method seems to underestimate the true number of classes (mode equal to one). 
	
	Finally, for Scenario III, the BIC in the joint latent class model seems to perform better than the proposed approach where it always selects the correct number of classes (one). This is not surprising, since the BIC always underestimated the true number of classes in the previous Scenarios. Using the proposed approach and assuming that the $\psi$ is equal to 15\%, we obtain 43\% of the time the correct number of class and a mode equal to two. In this Scenario some convergence problems were detected. When recalculating the \% and mode including only the simulations that were converged we obtain similar percentages for the correct number of class and a mode equal to 1 when $\psi$ is 15\%.

	\begin{table}[h!]
		\caption{Simulation results: Cut off $\psi$, percentage of true number of classes, mode of the number of classes}
		\centering
		\begin{tabular}{lrrr}
			\hline
			& $\psi$ (\%) & true \# of classes (\%) & mode of \# of classes \\
			\hline 
			\it{Scenario I: }  &&&\\
			\it{150 simulations}  &&&\\
			\hline 
			&1  & 0 & 6   \\ 
			&2  & 1 & 6  \\ 
			&5  & 23 & 5   \\ 
			&8  & 27 & 4   \\
			&10  & 39 & 4  \\
			&12  & 56 & 3  \\
			&15  & 55 & 3   \\
			\hline 
			\it{Scenario II: } &&&\\
			\it{150 simulations} &&&\\
			\hline 
			&1  & 0 & 6   \\ 
			&2  & 2 & 5 \\ 
			&5  & 31 & 3   \\ 
			&8  & 47 & 2   \\
			&10 & 51 & 2  \\
			&12  & 53 & 2  \\
			&15  & 57 & 2   \\
			\hline
			\it{Scenario III: } &&&\\
			\it{150 simulations} &&&\\
			\hline 
			&1  & 0 & 6   \\ 
			&2  & 0 & 6 \\ 
			&5  & 0 & 4   \\ 
			&8  & 11 & 2   \\
			&10 & 23 & 2  \\
			&12  & 34 & 2  \\
			&15  & 43 & 2   \\
		\end{tabular}
		\label{SimRes}
	\end{table}

	\section{Discussion}\label{Dis}
	In this paper we proposed a shared parameter joint model incorporating latent classes. Applying it to CF data, this model accounted for patient heterogeneity inherent in the progression of $FEV_1$. Compared to previously proposed joint latent class models \citep{proust2014joint} we obtained the strength of the association between $FEV_1$ and time-to first exacerbation per group of patients. Finally, we focused on the selection of the optimal number of classes and used an overfitted mixture model (high number of classes) to obtain the non-empty classes.
	
	A limitation of this approach is that it requires an intensive computational effort. In particular, for the class selection, where a model with a high number of classes is required, the number of parameters increases drastically. This, in combination with the high number of observations in the CF application increases the computational time that is required. Considering the difficulty of this model, it is almost impossible to obtain the optimal number of classes with other Bayesian criterion. Implementing the proposed criterion is straightforward; however, due to the complexity of the model it is computationally expensive to fit a model with a larger number of classes, e.g., 10. This could also explain the fact the a higher percentage for the predefined number $\psi$ was required in order to obtain the non-empty number of classes. It was shown in the simulation analysis that the BIC always underestimated the true number of parameters and it, therefore, performed better when the true number of classes was one. Even though, in that Scenario the proposed method did not work perfectly, it seems to be better than other criteria and easier to perform.
	
	Although there is a large database available in the US Registry, we used only a subset in order to make it feasible to run the proposed model. This subset has particular characteristics and it cannot be generalized to all patients in the Registry. Therefore, the presented results do not reflect the diversity of the whole database.
	
	Possible extensions would be to include more covariates also in the survival submodel in order to take into account extra information regarding the patients. Furthermore, using the proposed model for obtaining future $FEV_1$ measurement and time-to first exacerbation probabilities, could lead to more efficient treatment prioritization and clinical management for patients with CF.

	\bibliographystyle{abbrvnat} 
	\bibliography{mybiblo}

	\newpage
	
    \section*{Appendix}
\renewcommand{\thetable}{A\arabic{table}}

\setcounter{table}{0}

	\begin{table}[ht]
		\caption{Mean (standard deviation) of age at baseline, $FEV_1$ at baseline and number of visits at last follow-up visit per class.}
		\centering
		\begin{tabular}{rlll}
			\hline
			& Age at baseline & $FEV_1$ at baseline &Numvisityr \\ 
			&&&(Number of visits at the last \\
			&&&follow-up within the prior year)\\ 
			\hline
			Class 1 & 15 (4) & 59 (23) & 5 (3)\\ 
			Class 2 & 15 (4) & 64 (16) & 4 (3)\\ 
			Class 3 & 14 (5) & 46 (16) & 5 (3)\\ 
			\hline
		\end{tabular}
		\label{ClassesCont}
	\end{table}
	
	\newpage
	
	\begin{table}[ht]
		{\footnotesize
			\caption{Percentage of categorical varibles at baseline per class.}
			\centering
			\begin{tabular}{rrrrr}
				\hline
				\multicolumn{5}{l}{Gender:} \\
				& Females & Males &&\\ 
				\hline
				Class 1 & 0.32 & 0.23  &&\\ 
				Class 2 & 0.09 & 0.09  &&\\ 
				Class 3 & 0.16 & 0.11  &&\\ 
				\hline\hline
				\multicolumn{5}{l}{Number of F508del  alleles (genotype):} \\
				& Homozygous & Heterozygous & Neither & Missing \\ 
				\hline
				Class 1 & 0.27 & 0.21 & 0.03 & 0.04 \\ 
				Class 2 & 0.12 & 0.03 & 0.02 & 0.01 \\ 
				Class 3 & 0.14 & 0.08 & 0.02 & 0.02 \\ 		
				\hline\hline	
				\multicolumn{5}{l}{Hispanic} \\
				& No & Yes &&\\ 
				\hline
				Class 1 & 0.52 & 0.03 &&\\ 
				Class 2 & 0.16 & 0.02 &&\\ 
				Class 3 & 0.24 & 0.03 &&\\ 
				\hline\hline
				\multicolumn{5}{l}{White} \\
				& No & Yes &&\\ 
				\hline
				Class 1 & 0.01 & 0.54 &&\\ 
				Class 2 & 0.00 & 0.18 &&\\ 
				Class 3 & 0.01 & 0.26 &&\\ 
				\hline\hline
				\multicolumn{5}{l}{SESlow (Using state/federal or having no insurance is a marker of low socioeconomic status):}\\
				& No & Yes &&\\ 
				\hline
				Class 1 & 0.31 & 0.24  &&\\ 
				Class 2 & 0.09 & 0.09  &&\\ 
				Class 3 & 0.12 & 0.15  &&\\ 
				\hline\hline
				\multicolumn{5}{l}{MRSA (Methicillin-resistant Staphylococcus aureus):} \\
				& No & Yes &&\\ 
				\hline
				Class 1 & 0.51 & 0.04 &&\\ 
				Class 2 & 0.17 & 0.02 &&\\ 
				Class 3 & 0.24 & 0.03 &&\\ 
				\hline\hline
			    \multicolumn{5}{l}{MSSA (Methicillin-sensitive Staphylococcus aureus):} \\
				& No & Yes &&\\ 
				\hline
				Class 1 & 0.43 & 0.12 &&\\ 
				Class 2 & 0.14 & 0.04 &&\\ 
				Class 3 & 0.22 & 0.05 &&\\ 		
				\hline\hline
				\multicolumn{5}{l}{Pa (Pseudomonas aeruginosa):} \\
				& No & Yes &&\\ 
				\hline
				Class 1 & 0.32 & 0.23 &&\\ 
				Class 2 & 0.11 & 0.07 &&\\ 
				Class 3 & 0.17 & 0.10 &&\\ 
				\hline\hline
				\multicolumn{5}{l}{Aspergillus:} \\
				& No & Yes &&\\ 
				\hline
				Class 1 & 0.42 & 0.13 &&\\ 
				Class 2 & 0.14 & 0.04 &&\\ 
				Class 3 & 0.20 & 0.07 &&\\ 
				\hline\hline
				\multicolumn{5}{l}{CFRD (CF-related diabetes):} \\	
				& Normal & Impaired & CFRD with or without &\\ 
				&&& fasting hyperglycemia & \\
				\hline
				Class 1 & 0.47 & 0.02 & 0.07 &\\ 
				Class 2 & 0.15 & 0.01 & 0.02 &\\ 
				Class 3 & 0.23 & 0.01 & 0.03 &\\ 
				\hline\hline
				\multicolumn{5}{l}{PancEnzymes (Taking a pancreatic enzyme supplement, marks pancreatic insufficiency):} \\
				& No & Yes &&\\ 
				\hline
				Class 1 & 0.49 & 0.06 &&\\ 
				Class 2 & 0.17 & 0.02 &&\\ 
				Class 3 & 0.24 & 0.03 &&\\ 	
			\end{tabular}
		}
		\label{ClassesCat}
	\end{table}

\end{document}